\documentclass{article}
\usepackage[utf8]{inputenc}

\usepackage{verbatim}

\usepackage{amssymb}
\usepackage{amsmath}
\usepackage{mathrsfs}
\usepackage{mathtools}
\usepackage{color}

\title{Lie point symmetries of near-horizon geometry equation}
\author{E. Buk, J. Lewandowski, A. Szereszewski}
\date{\small \emph{Faculty of Physics, University of Warsaw,\\ Pasteura 5, 02-093 Warsaw, Poland}}

\begin{document}

\maketitle

\begin{abstract} All the Lie point symmetries of the near extremal horizon geometry equation, in the case of $4$-dimensional Einstein vacuum spacetime with cosmological constant, are the diffeomorphisms of the space of the null generators of the horizon. This result is also generalised to the Maxwell-Einstein spacetime.    
\end{abstract}
\section{Introduction}
Subject to the near-horizon geometry (NHG) equation is a pair: a metric tensor of the Riemannian signature and a differential $1$-form field. They are defined on any $n$ dimensional manifold. This equation is satisfied by the metric tensor and rotation $1$-form potential induced on a spacelike section of an extremal isolated horizon (for example a degenerate Killing horizon) in $(n+2)$-dimensional Einstein vacuum spacetime with a cosmological constant \cite{H}--\cite{LP2005}. The equation is also satisfied by a metric tensor and {\it minus} rotation $1$-form potential induced on a spacelike slice of every leaf of a non-expanding null surface foliation of Kundt's vacuum spacetime with a cosmological constant (that coincidence is still somewhat mysterious, although it has been partially understood \cite{PLJ}--\cite{PO}). 
 
The NHG equation attracts attention of theoretical physicists and mathematical relativists for several reasons. The first one is that the NHG equation implies local black hole uniqueness results \cite{LP2003}--\cite{KLS}. An open problem at this point is the existence of non-axisymmetric solutions on $S_2$. If it were proved they do not exist, that would help in completing the global black hole uniqueness theorems \cite{CCH}. The non-existence of non-trivial non-rotating solutions \cite{CRT} and non-existence of rotating solutions on positive genus $2$-manifolds \cite{DRKLS} also have counterparts in the global BH theory. The second reason is the relation with the so called near-horizon geometries -- exact solutions to the Einstein equations that can be obtained via the Geroch-Bardeen-Horowitz limit \cite{BH} from spacetimes containing a degenerate Killing horizon. They are used to define extremal quantum black hole entropy \cite{S} and to introduce the extremal isolated horizon/CFT correspondence \cite{WT}. The third reason is that a new solution to the NHG equation would give us a new exact solution to the Einstein equations \cite{PLJ}. 
 
The NHG equation was also generalised to the Einstein-Maxwell, Einstein-Yang Mills and other matter fields cases \cite{LP2003,KL1,DGS}. 
 
Despite the extensive literature on the NHG equation, there are still open problems concerning it. In this paper we solve one of them, namely finding its Lie point symmetries. We also consider and solve that problem for Maxwell-NHG equation.
 
\section{The near-horizon geometry equation}
Consider a $2$ dimensional manifold $S$ endowed with a Riemannian metric tensor~$g^{(2)}$,
\begin{equation} \label{g}
	g^{(2)}=g_{AB}dx^Adx^B,
\end{equation}
 and a differential 1-form field $\omega$,
\begin{equation} \label{omega} 
	\omega = \omega_A dx^A, 
\end{equation}
where the indices $A,B$ pertain to local coordinates $(x^A)=(x^1,x^2)$ defined on $S$. 
The near-horizon geometry equation with a cosmological constant $\Lambda$ reads
\begin{equation} \label{NHG0}
	\nabla_{(A}\omega_{B)}
		+\omega_{A}\omega_{B}-\frac{1}{2}R_{AB} 
		+\frac{1}{2}\Lambda g_{AB} =0
\end{equation}
where $\nabla$ is the torsion free covariant derivative compatible with $g^{(2)}$ and $R_{AB}dx^{A}dx^{B}$ is its Ricci tensor. There exist (locally) conformally flat coordinates $(x^1,x^2)$, that is coordinates such that
\begin{equation} \label{Re_met}
	g^{(2)}=e^{\phi}\left[\big(dx^1\big)^2+\big(dx^2\big)^2\right], 
\end{equation}
where $\phi=\phi(x^1,x^2)$. We will be using them throughout this paper. Now, the Ricci tensor featuring in the equation takes the following form
$$R_{AB}dx^Adx^B = -\frac{1}{2}\big(\phi_{,11}+\phi_{,22}\big)e^{-\phi}g^{(2)}.$$ Upon the decomposition in the coordinates (\ref{Re_met}) the NHG equation (\ref{NHG0}) amounts to a system of the following 3 equations:
\begin{equation} \label{NHG_real}
	\left\{
	\begin{aligned}
			2\omega_{1,1}
			- \omega_{1}\phi_{,1} 
				+\omega_{2}\phi_{,2} 
			+2 \omega_{1}^2 + 
			\frac{1}{2}\left(
				\phi_{,11} 
				+\phi_{,22}
			\right)
			+\Lambda e^{\phi}
			&= 0
		\\
			2\omega_{2,2}
			+ 
				\omega_{1}\phi_{,1} 
				-\omega_{2}\phi_{,2}
			+2\omega_{2}^2 +
			\frac{1}{2}\left(
				\phi_{,11} 
				+\phi_{,22}
			\right)
			+\Lambda e^{\phi}
			&= 0
		\\
		%
				\omega_{1,2} 
				+\omega_{2,1}
			- 
				\omega_{1}\phi_{,2} 
				-\omega_{2}\phi_{,1}
			+2\omega_{1}\omega_{2} 
			&= 0
	\end{aligned}
	\right.
\end{equation}

\section{Symmetries of near-horizon geometry equation}
A symmetry of the system of the equations (\ref{NHG_real}) is called a Lie point symmetry whenever it maps a quintuple $(x^1,x^2,\omega_1,\omega_2,\phi)$ into quintuple $(\tilde{x}^1,\tilde{x}^2,\tilde{\omega}_1,\tilde{\omega}_2,\tilde{\phi})$ of functions defined on $S$ such that at every point $p\in S$, the values of $\tilde{x}^1, \tilde{x}^2, \tilde{\omega}_1, \tilde{\omega}_2, \tilde{\phi}$ are determined by the values of $x^1,x^2,\omega_1(x^1,x^2),\omega_2(x^1,x^2),\phi(x^1,x^2)$. A Lie point symmetry we seek in this paper is a $1$-dimensional group of point symmetries of the equations (\ref{NHG_real}). It can be generated by a vector field 
\begin{equation}\label{X_0}
	X_0 = 
		\xi^{A}(x^{B}, \omega_B,\phi)\frac{\partial}{\partial x^{A}} + 
		\eta_A(x^{B}, \omega_B,\phi)\frac{\partial}{\partial \omega_A} + \zeta(x^{B}, \omega_B,\phi)\frac{\partial}{\partial \phi}
\end{equation}
 where the values of the functions $x^B, \omega_A(x^B), \phi(x^B)$ are considered as independent variables.
 
Let us introduce for the equations (\ref{NHG_real}) a short notation
\begin{equation}\label{H_k0}
	H_k(\omega_A,\phi, \omega_{A,B},\phi_{,A}, \phi_{,AB} ) = 0,
	\qquad
	k=1,2,3,
\end{equation}
where $H_1, H_2$ and $H_3$ are the left hand sides of the equations (\ref{NHG_real}). Given a vector field $X_0$ (\ref{X_0}), the infinitesimal action of its flow on the equation (\ref{NHG_real}) amounts to the action on the functions $H_k$ with a suitable prolongation of $X_0$, namely with the following vector field
\begin{equation}
	X =
		\xi^{A}\frac{\partial}{\partial x^{A}} + \eta_A\frac{\partial}{\partial \omega_A} + \zeta \frac{\partial}{\partial \phi} +
		\eta_{A B}\frac{\partial}{\partial \omega_{A ,B}} + \zeta_A \frac{\partial}{\partial \phi_{,A}} +\zeta_{AB} \frac{\partial}{\partial \phi_{,AB}},
		\end{equation}
where the coefficients $\xi^A$, $\eta_A$ and $\zeta$ are the same as in (\ref{X_0}), while $\eta_{AB}, \, \zeta_A$ and $\zeta_{AB}$ can be expressed in terms of derivatives of the functions $\xi^A$, $\eta_A$ and $\zeta$ according to the following formulas:
\begin{align}
	\eta_{AB} &= \mathcal{D}_B\big(\eta_A -\xi^C\omega_{A,C}\big)+\xi^C\omega_{A,BC},\label{eq1:higher_etas}
	\\
	\zeta_{A} &=\mathcal{D}_A\big(\zeta -\xi^B\phi_{,B}\big)+\xi^B\phi_{,BA},\label{eq2:higher_etas}
	\\
	\zeta_{AB} &= \mathcal{D}_B\mathcal{D}_A\big(\zeta -\xi^C\phi_{,C}\big)+\xi^C\phi_{,CAB}. \label{eq3:higher_etas}
\end{align}		
Operator $\mathcal{D}_A$ used in (\ref{eq1:higher_etas})--(\ref{eq3:higher_etas}) is a total derivative defined for any function $Q(x^A,\omega_A,\phi,\omega_{A,B},\phi_{,A},\phi_{AB})$ in the following way
\begin{equation}
	\mathcal{D}_A Q=\frac{\partial Q}{\partial x^A}+\omega_{B,A}
	\frac{\partial Q}{\partial \omega_B}+\phi_{,A}\frac{\partial Q}{\partial \phi}+\omega_{B,CA}
	\frac{\partial Q}{\partial \omega_{B,C}}+\phi_{,BA}\frac{\partial Q}{\partial \phi_{,B}}+\phi_{,BCA}\frac{\partial Q}{\partial \phi_{,BC}}.
\end{equation}

The vector field $X_0$ generates a symmetry of the equation (\ref{NHG_real}) if and only if the vector field $X$ is tangent to the surfaces defined by (\ref{H_k0}), that is iff \cite{olver_lie}:
\begin{equation} \label{eq:symmetr_eq}
	X(H_{k})_{|_{H_1,H_2,H_3=0}} = 0, \qquad k=1,2,3.	
\end{equation}
	 
Our aim now, is to find all the vector fields $X_0$ that satisfy the conditions (\ref{eq:symmetr_eq}). The conditions amount to a system of first order linear partial differential equations for the functions $\xi^{A}(x^B,\omega_B,\phi)$, $\eta_{A}(x^B,\omega_B,\phi)$ and $\zeta(x^B,\omega_B,\phi)$ obtained in three steps. The first one is the insertion of the functions $H_1,H_2, H_3$ given by the left hand sides of the NHG equations (\ref{NHG_real}) into (\ref{eq:symmetr_eq}), resulting in:
\begin{equation}\label{XH}
	\left\{
	\begin{aligned}
		(\phi_{,1}-4\omega_1)\eta_1-\phi_{,2}\eta_2+\Lambda e^{\phi}\zeta 
		 -2\eta_{11}-\omega_1\zeta_1+\omega_2\zeta_2+\frac{1}{2}\big(\zeta_{11}+\zeta_{22}\big) &= 0 \\
		(\phi_{,2}-4\omega_2)\eta_2-\phi_{,1}\eta_1+\Lambda e^{\phi}\zeta
		 -2\eta_{22}+\omega_1\zeta_1-\omega_2\zeta_2+\frac{1}{2}\big(\zeta_{11}+\zeta_{22}\big) &= 0 \\
		(\phi_{,2}-2\omega_2)\eta_1+(\phi_{,1}-2\omega_1)\eta_2-\eta_{12}-\eta_{21}-\omega_2\zeta_1 -\omega_1\zeta_2 &= 0
	\end{aligned}
	\right.
	.
\end{equation}	
The second step is eliminating the functions $\eta_{AB}, \zeta_A$ and $\zeta_{AB}$ in favour of the unknowns $\xi^A,\eta_A,\zeta$ using the equations (\ref{eq1:higher_etas})--(\ref{eq3:higher_etas}). The third step is choosing among the variables $(\phi, \phi_{,A},\phi_{,AB}, \omega_A,\omega_{A,B})$ a maximal set of independent ones parametrising freely the surface $H_k=0$, $k=1,2,3$, and expressing by them the remaining three dependent variables. Then, the left hand sides of (\ref{XH}) take the form of three polynomials in the independent variables. They vanish if and only iff every coefficient vanishes independently.

It turns out that the conditions on the functions $\xi^{A}$ boil down to the Cauchy-Riemann equations 
\begin{equation} \label{eq:xi_constraint}
	\xi^{1}_{\ ,1} = \xi^{2}_{\ ,2}
	\qquad
	\xi^{1}_{\ ,2} = -\xi^{2}_{\ ,1} 
\end{equation}	
and that:
\begin{equation} \label{eq:constraint_after_xi_constraint}
	\frac{\partial\zeta}{\partial \omega_{1}} = 
	\frac{\partial\zeta}{\partial \omega_{2}} = 0.
\end{equation}
This information simplifies formulas (\ref{eq1:higher_etas})--(\ref{eq3:higher_etas}) considerably.
Finally, we obtain a general solution to the Lie point symmetry equations, the vector field $X_0$ in the form:
\begin{equation} \label{VecField1}
	\begin{split}
	X_0 =\ &
		\xi^{1}\frac{\partial}{\partial x^{1}}
		+\xi^{2}\frac{\partial}{\partial x^{2}}
		-\Big(
			\xi^{1}_{\ ,1}\omega_{1} 
			-\xi^{1}_{\ ,2}\omega_{2} 
		\Big) \frac{\partial}{\partial \omega_{1}}
		-\Big(
			\xi^{1}_{\ ,2}\omega_{1} 
			+\xi^{1}_{\ ,1}\omega_{2} 
		\Big) \frac{\partial}{\partial \omega_{2}} \\
		& -2\Big(
			\xi^{1}_{\ ,1}
			+l
		\Big)\frac{\partial}{\partial \phi}\ ,
	\end{split}
\end{equation}
where functions $\xi^A(x^1,x^2)$ satisfy (\ref{eq:xi_constraint}) and $l$ is a constant, which is non-zero only if $\Lambda = 0$. Part of this generator can be integrated to obtain the finite transformation itself (see first five formulas of (\ref{symm_tr_real}) in Appendix).

Interpretation of the derived symmetries is easy. The part $\xi^1\frac{\partial}{\partial x^1}+\xi^2\frac{\partial}{\partial x^2}$ of the vector field $X_0$ generates conformal transformations preserving the form (\ref{Re_met}) of the metric tensor - this is the meaning of the Cauchy-Riemann equations. Those transformations generate transformations of the $1$-form field $\omega_A$ and the conformal factor $\phi$. They correspond to the remaining part of the vector field $X_0$. Our result can be concluded as follows:

\medskip\noindent
\textbf{Theorem 1.} \textit{All the  Lie point symmetries of the equation (\ref{NHG_real}) given by the NHG equation (\ref{NHG0}) written on a $2$-manifold $S$ in  conformally flat coordinates (\ref{Re_met}) are the diffeomorphisms of $S$  preserving the conformal structure.}
\medskip

One can also consider the Lie point symmetries of the NHG equation (\ref{NHG0}) written in general coordinates. But  every point symmetry of (\ref{NHG0}) can be mapped by some coordinate transformations  into a point symmetry of  (\ref{NHG_real}). In that way Theorem 1 provides also all the Lie point transformations of the equation (\ref{NHG0}). 

\medskip\noindent
\textbf{Corollary 1.} \textit{All the Lie point symmetries of the equation (\ref{NHG0})  defined on a manifold $S$ are  the diffeomorphisms of $S$.}

\section{The NHG equation coupled to the Maxwell equations}
The NHG equation is generalised to the Einstein-Maxwell equations. In the case of $4$ dimensional spacetimes the resulting system of equations is defined again on a $2$ dimensional manifold $S$ in terms of a Riemannian metric tensor $g^{(2)}$ (\ref{g}) and a differential 1-form field $\omega$ (\ref{omega}). A new element is a complex valued function 
 $\Phi_1$ defined on $S$ representing the Maxwell field. To spell out the equations, it is convenient to introduce a null frame 
$$ g^{(2)}_{AB}= m_A\bar{m}_B + m_B\bar{m}_A.$$ 

The NHG-Maxwel equations are the following system
\begin{align}
	&\nabla_{(A}\omega_{B)}
		+\omega_{A}\omega_{B}-\frac{1}{2}R_{AB} 
		+\frac{1}{2}\Lambda g_{AB}
		+2\left|\Phi_1\right|^{2} g_{AB}
		= 0,	
		\label{NHG1}\\
	&\bar{m}^A \left( \partial_A+2\omega_A \right)\Phi_{1}
		=0, 
		\label{MaxEqn}
\end{align}
where $R_{AB}$ and $\nabla_A$ are the same as in the previous section. We split $\Phi_1$ into the real and imaginary part, 
$$\Phi_1=U+iV,$$ 
and use the conformally flat coordinates (\ref{Re_met}). The resulting equations (\ref{NHG1})-(\ref{MaxEqn}) read as follows:
\begin{equation} \label{MNHG_real}
	\left\{
	\begin{aligned}
			2\omega_{1,1}
			- \omega_{1}\phi_{,1} 
			+\omega_{2}\phi_{,2} 
			+2\omega_{1}^2
			+\frac{1}{2}\left(
				\phi_{,11} 
				+\phi_{,22}
			\right)
			+\Lambda e^{\phi}
			+2\left(
				U^{2}
				+V^{2}
			\right) e^{\phi}
			&= 0
		\\
			2\omega_{2,2}
			+\omega_{1}\phi_{,1} 
			-\omega_{2}\phi_{,2}
			+2\omega_{2}^2
			+\frac{1}{2}\left(
				\phi_{,11} 
				+\phi_{,22}
			\right)
			+\Lambda e^{\phi}
			+2\left(
			 	U^{2}
			 	+V^{2}
			\right) e^{\phi} 
			&= 0
		\\
			\omega_{1,2} 
			+\omega_{2,1}
			-\omega_{1}\phi_{,2} 
			-\omega_{2}\phi_{,1}
			+2\omega_{1}\omega_{2} 
			&= 0
		\\
			U_{,1} 
			-V_{,2} 
			+2\left( 
				\omega_{1}U
				-\omega_{2}V
			\right) 
			&= 0
		\\
			U_{,2}
			+V_{,1} 
			+2\left( 
				\omega_{2}U
				+\omega_{1}V
			\right) 
			&= 0 
	\end{aligned}
	\right.
	.
\end{equation} 

The analogues symmetry analysis of the system (\ref{MNHG_real}) leads to the following symmetry generating vector field:
\begin{equation} \label{VecField2}
	\begin{split}
	X_0 =\ &
		\xi^{1}\frac{\partial}{\partial x^{1}}
		+\xi^{2}\frac{\partial}{\partial x^{2}}
		-\Big(
			\xi^{1}_{\ ,1}\omega_{1} 
			-\xi^{1}_{\ ,2}\omega_{2} 
		\Big) \frac{\partial}{\partial \omega_{1}}
		-\Big(
			\xi^{1}_{\ ,2}\omega_{1} 
			+\xi^{1}_{\ ,1}\omega_{2} 
		\Big) \frac{\partial}{\partial \omega_{2}} \\
		& -2\Big(
			\xi^{1}_{\ ,1}
			+l
		\Big)\frac{\partial}{\partial \phi}
		+\Big(
			l U
			-a V
		\Big) \frac{\partial}{\partial U}
		+\Big(
			a U
			+l V
		\Big) \frac{\partial}{\partial V}
		.
	\end{split}
\end{equation}
As before, functions $\xi^A(x^1,x^2)$ satisfy (\ref{eq:xi_constraint}), while $l$ and $a$ are constants, wherein the former is non-zero only if $\Lambda = 0$. This result can be encapsulated in the following:
\medskip\noindent

\textbf{Theorem 2.} \textit{All the Lie point symmetries of the Maxwell--NHG equations (\ref{MNHG_real}) on a $2$-manifold $S$ are those induced by the diffeomorphisms of $S$ preserving the form (\ref{Re_met}) of the metric tensor supplemented by the following global transformation of electromagnetic field $\Phi_1 \mapsto e^{i\alpha}\Phi_1$, where 
$\alpha\in\mathbb{R}$.}
\medskip\noindent

Using the same arguments as in the previous section we conclude that:

\medskip\noindent
\textbf{Corollary 2.} \textit{All the Lie point symmetries of the system of Maxwell--NHG equations (\ref{NHG1})--(\ref{MaxEqn})  defined on a $2$ dimensional manifold $S$ are  the diffeomorphisms of $S$.}

\section{Summary} 
We have found all the Lie point symmetries of the near-horizon geometry equation in the vacuum and electro-vacuum case, respectively.
Our results are presented in Theorem 1, Corollary 1, Theorem 2 and Corollary 2. They are  summarised  below.

The only  Lie point symmetries of the NHG equation (\ref{NHG0}) written on a $2$-manifold $S$ in a conformally flat coordinates (\ref{Re_met}) are those induced by the diffeomorphisms of $S$ preserving the conformal structure. They are generated by the vector field (\ref{VecField1}).  In a consequence,   the only symmetries  the NHG equation in arbitrary coordinates are the diffeomorphisms of $S$. 

For the Maxwell--NHG equations (\ref{NHG1})--(\ref{MaxEqn}) on $S$ in the conformally flat coordinates, the only Lie point symmetries are again, those induced by the diffeomorphisms of $S$ preserving the form (\ref{Re_met}) of the metric tensor, supplemented by the global rescaling of electromagnetic field by a constant. Generator of these symmetries is given by (\ref{VecField2}).  Relaxing the coordinates to arbitrary systems results in admitting all the diffeomorphisms of $S$ as Lie point symmetries. 

\bigskip

\noindent{\bf Acknowledgements} 

\bigskip 
\noindent
JL thanks Michal Grundland for suggesting the investigation of the Lie point symmetries of the NHG equation. The work was supported by the grant  Opus number 2017/27/B/ST2/02806 of the Polish  National Science Center.  

\section*{Appendix: Calculation of symmetries}	
Here we shall discuss the derivation of the symmetry vector field (\ref{VecField2}) in more detail and in addition present the symmetry transformation it induces. 

For convenience, we denote
\begin{equation}
	u^1 = \omega_1, \qquad u^2= \omega_2, \qquad u^3=\phi, \qquad u^4= U,\qquad u^5 = V
\end{equation}	
and write symmetry vector filed in the following form ($A,B,C=1,2$, $\alpha,\beta=1,2,\dots,5$)
\begin{equation}
	X =	\xi^{A}(x^{C}, u^{\beta})\frac{\partial}{\partial x^{A}} + 
		\eta^{\alpha}(x^{C}, u^{\beta})\frac{\partial}{\partial u^{\alpha}} + 
		\eta^{\alpha}_{\ A}(x^{C}, u^{\beta})\frac{\partial}{\partial u^{\alpha}_{\ ,A}} + \eta^{\alpha}_{\ AB}(x^{C}, u^{\beta})\frac{\partial}{\partial u^{\alpha}_{\ ,AB}},
\end{equation}	
where $\eta^{\alpha}_{\ A}$ and $\eta^{\alpha}_{\ AB}$ can be expressed in terms of derivatives of functions $\xi^{A}(x^{B}, u^{\beta})$ and $\eta^{\alpha}(x^{A}, u^{\beta})$ according to
\begin{equation} \label{etas:u}
	\begin{split}
		\eta^{\alpha}_{\ A} =\ &
			\eta^{\alpha}_{\ ,A}
			+\frac{\partial \eta^{\alpha}}{\partial u^{\beta}}u^{\beta}_{\ ,A}
			-\xi^{B}_{\ ,A}u^{\alpha}_{\ ,B}
			-\frac{\partial \xi^{B}}{\partial u^{\beta}}u^{\beta}_{\ ,A}u^{\alpha}_{\ ,B},
		\\
		\eta^{\alpha}_{\ AB} =\ &
			\eta^{\alpha}_{\ ,AB}
			+\frac{\partial \eta^{\alpha}_{\ ,A}}{\partial u^{\beta}}u^{\beta}_{\ ,B}
			+\frac{\partial \eta^{\alpha}_{\ ,B}}{\partial u^{\beta}}u^{\beta}_{\ ,A}
			-\xi^{C}_{\ ,AB}u^{\alpha}_{\ ,C}
			+\frac{\partial^2 \eta^{\alpha}}{\partial u^{\beta}\partial u^{\gamma}}u^{\beta}_{\ ,A}u^{\gamma}_{\ ,B} 
		\\
			&-\left( 
				\frac{\partial \xi^{C}_{\ ,A}}{\partial u^{\beta}}u^{\beta}_{\ ,B}
				+\frac{\partial \xi^{C}_{\ ,B}}{\partial u^{\beta}}u^{\beta}_{\ ,A}
			\right)	u^{\alpha}_{\ ,C}
			-\frac{\partial^2 \xi^{C}}{\partial u^{\beta}\partial u^{\gamma}}u^{\beta}_{\ ,A}u^{\gamma}_{\ ,B}u^{\alpha}_{\ ,C} 
			+\frac{\partial \eta^{\alpha}}{\partial u^{\beta}}u^{\beta}_{\ ,AB} 
		\\
			&-\xi^{C}_{\ ,A}u^{\alpha}_{\ ,BC}
			-\xi^{C}_{\ ,m}u^{\alpha}_{\ ,AC}
			-\frac{\partial \xi^{C}}{\partial u^{\beta}}
			\left(
				u^{\alpha}_{\ ,C}u^{\beta}_{\ ,AB}
				+u^{\beta}_{\ ,A}u^{\alpha}_{\ ,BC}
				+u^{\alpha}_{\ ,AC}u^{\beta}_{\ ,B}
			\right).
	\end{split}
\end{equation}
By rewriting the system (\ref{MNHG_real}) as $H_k(x^A,u^\alpha)=0$ ($k=1,2,\dots,5$), the symmetry conditions \cite{olver_lie} read
\begin{equation} \label{MNHGeq:symmetr_eq}
	X(H_{k}) = 0, \quad k=1,\dots,5 \qquad \text{whenever}\qquad H_1=H_2=\dots=H_5=0		,
\end{equation}
which induce the system of first order partial differential equations for both $\xi^{A}(x^B,u^\beta)$ and $\eta^{\alpha}(x^B,u^\beta)$. Once these functions are found, the flow generated by 
\begin{equation}
	X_0 = 
		\xi^{A}(x^{B}, u^{\beta})\frac{\partial}{\partial x^{A}} + 
		\eta^{\alpha}(x^{B}, u^{\beta})\frac{\partial}{\partial u^{\alpha}}
\end{equation}
gives rise to the symmetries 
\begin{equation}
 \tilde{x}^A=\tilde{x}^A(x^B,u^\beta), \qquad 
 \tilde{u}^\alpha=\tilde{u}^\alpha(x^B,u^\beta)
	\end{equation}	
of system (\ref{MNHG_real}).	
Insertion of Maxwell-NHG equations (\ref{MNHG_real}) into (\ref{MNHGeq:symmetr_eq}) produces the following system of partial differential equations for 7 functons $\xi^{A}(x^B,u^\beta)$ and $\eta^{\alpha}(x^B,u^\beta)$:
\begin{align}
	\left(\phi_{,1}-4\omega_1\right)\eta^1
		-\phi_{,2}\eta^2
		+\left(
			2(U^2+V^2)
			+\Lambda
		\right)e^{\phi}\eta^3
		+4 e^\phi\left(U\eta^4-V\eta^5\right)& \nonumber \\
	-2\eta^1\,_1
		-\omega_1\eta^3\,_1
		+\omega_2\eta^3\,_2
		+\frac{1}{2}\left(
			\eta^3\,_{11}
			+\eta^3\,_{22}
		\right)&=0 \label{SymEqns1}
	\\
	-\phi_{,1}\eta^1+
		\left(\phi_{,2}-4\omega_2\right)\eta^2
		+\left(
			2(U^2+V^2)
			+\Lambda
		\right)e^{\phi}\eta^3
		+4 e^\phi\left(U\eta^4-V\eta^5\right)& \nonumber \\
	-2\eta^2\,_2
		+\omega_1\eta^3\,_1
		-\omega_2\eta^3\,_2
		+\frac{1}{2}\left(
			\eta^3\,_{11}
			+\eta^3\,_{22}
		\right)&=0
	\label{SymEqns2}
	\\
	(\phi_{,2}-2\omega_2)\eta^1
		+(\phi_{,1}-2\omega_1)\eta^2
		-\eta^1\,_2
		-\eta^2\,_1
		-\omega_2\eta^3\,_1
		-\omega_1\eta^3\,_2&=0 \label{SymEqns3}
	\\
	2U\eta^1
	-2V\eta^2
		+2\omega_1\eta^4
		-2\omega_2\eta^5
		+\eta^4\,_1
		-\eta^5\,_2 &=0 \label{SymEqns4}
	\\
	2V\eta^1
		+2U\eta^2
		+2\omega_2\eta^4
		+2\omega_1\eta^5
		+\eta^4\,_2
		+\eta^5\,_1 &=0. \label{SymEqns5}
\end{align}		
To simplify calculations the following steps can be performed:
\begin{enumerate}
	\item 
		For each $\eta^{\alpha}_{\ AB}$ in (\ref{SymEqns1}) and (\ref{SymEqns2}) calculate terms containing second derivatives $u^{\alpha}_{\ ,AB}$ that do not appear in $H_{k} = 0$.
	\item
		Insert above mentioned terms into symmetry conditions (\ref{SymEqns1}) and (\ref{SymEqns2}). Application of $H_{k} = 0$ does not change these terms therefore they all must disappear.
	\item
		Calculate all terms containing second derivatives $u^{\alpha}_{\ ,AB}$ that appear in $\eta^{\alpha}_{\ AB}$ using previously obtained results. 
\end{enumerate}
This algorithm was outlined in \cite{stephani_symmetries}. After carrying out this routine it turns out that the functions $\xi^{A}(x^{B})$ obey the Cauchy-Riemann equations (\ref{eq:xi_constraint}) and they are independent of $u^\alpha$.
This information simplifies formulas (\ref{etas:u}) considerably. It is worth to note, that calculations for the first three equations (\ref{SymEqns1})--(\ref{SymEqns3}) can be now conducted independently from the last two (\ref{SymEqns4})--(\ref{SymEqns5}). Finally, we obtain the vector field $X_0$ given in (\ref{VecField2}) 
and simultaneously  prove Theorem 2. On the other hand, a derivation of vector field (\ref{VecField1}) which is equivalent to a proof of Theorem 1 was performed in a way analogous to the one outlined above,  in the vacuum case.

Integration of vector field (\ref{VecField2}) leads to the following symmetry transformation 	
\begin{equation}
	\begin{aligned}
		\frac{d}{dt} \tilde{x}^{1}(t) =\ &
			\xi^{1}\left( \tilde{x}^{1}(t), \tilde{x}^{2}(t) \right)
		\\
		\frac{d}{dt} \tilde{x}^{2}(t) =\ &
			\xi^{2}\left( \tilde{x}^{1}(t), \tilde{x}^{2}(t) \right)
		\\ 
		\tilde{\omega}_{1} =\ & 
			\omega_{1}e^{-\Xi_{1}}\cos \left( \Xi_{2} \right)  
			+\omega_{2}e^{-\Xi_{1}}\sin\left( \Xi_{2} \right)
		\\
		\tilde{\omega}_{2} =\ & 
			-\omega_{1}e^{-\Xi_{1}}\sin\left( \Xi_{2} \right) 
			+\omega_{2}e^{-\Xi_{1}}\cos\left( \Xi_{2} \right)
		\\
		\tilde{\phi} =\ &
			\phi -2\Xi_{1} - 2\lambda
		\\
		\tilde{U} =\ 
			&e^{\lambda}
			\left[
				U\cos\left( at \right) 
				+V\sin\left( at \right)
			\right]
		\\
		\tilde{V} =\ 
			&e^{\lambda}	
			\left[
				-U\sin\left( at \right) 
				+V\cos\left( at \right)
			\right]  \label{symm_tr_real}
	\end{aligned}
	,
\end{equation}
where $\lambda$ is constant (non-vanising only if $\Lambda = 0$), functions $\xi^A$ satisfy Cauchy-Riemann equations (\ref{eq:xi_constraint}) and
\begin{equation}  \label{Xis}
	\left\{
	\begin{aligned}
		\Xi_{1} =\ &
			\int_{0}^{t}\xi^{1}_{\ ,1}(s)ds
		\\
		\Xi_{2} =\ &
			\int_{0}^{t}\xi^{1}_{\ ,2}(s)ds
	\end{aligned}
	\right.  
	.
\end{equation}
The diffeomorphisms of independent variables must be left in implicit form, as symmetry conditions do not allow for further specification.The above result is complicated by the fact that we set $g^{(2)}$ in conformally flat form and used real coordinates. This difficulty can be circumvented by introduction of complex coordinates on $S$. 
Hence if one introduce the complex coordinate
$z = (x^{1}+ix^{2})/\sqrt{2}$
and function
\begin{equation}
	\omega = \frac{1}{\sqrt{2}} \left(\omega_{1}-i\omega_{2}\right),
\end{equation}
then the metric (\ref{Re_met}) becomes
\begin{equation}
	g^{(2)} = 2e^{\phi}dzd\bar{z}
\end{equation}
and Maxwell-NHG system (\ref{MNHG_real}) reads
\begin{equation} \label{MNHG_complex}
	\begin{aligned}
		\omega_{,z}-\phi_{,z}\omega +\omega^{2} =\ & 0{}
		\\
		\omega_{,\bar{z}} 
		+\bar{\omega}_{,z} 
		+2\omega\bar{\omega} 
		+\phi_{,z\bar{z}} 
		+\Lambda e^{\phi}
		+2 e^{\phi} \left| \Phi_{1} \right|^{2}
			=\ & 0
		\\
		(\Phi_{1})_{,\bar{z}} + 2\bar{\omega}\Phi_{1} =\ & 0
	\end{aligned}
	\quad, 
\end{equation} 
where functions $\omega (z, \bar{z})$ and $\Phi_{1} (z, \bar{z})$ are complex, and $\phi (z, \bar{z})$ is real.
The symmetries of (\ref{MNHG_complex}) yield 
\begin{equation}
 \tilde{z} = f(z), \qquad \tilde{\omega}=\omega/f'(z), \qquad \tilde{\phi}=\phi-\log|f'\bar{f}'|-2\lambda, \qquad \tilde{\Phi}_1=e^{\lambda+i\alpha}\Phi_1, \label{tr_complex}
 \end{equation}
where $f(z)$ is arbitrary non-constant holomorphic function and $\lambda$, $\alpha$ are real constants.
It is noted that for non-vanishing $\lambda$ the system (\ref{MNHG_complex}) is transformed under (\ref{tr_complex}) to the same one with cosmological constant $\Lambda$ replaced by $\tilde{\Lambda}=\Lambda e^{2\lambda}$.
 
For completeness, we also present the form of the vector field which generates symmetry transformation (\ref{tr_complex}). Using complex coordinates $X_0$ reads 
\begin{equation}
\begin{aligned}
	X_0 =\ 
		\xi\frac{\partial}{\partial z}
		+\bar{\xi}\frac{\partial}{\partial \bar{z}}
		-\xi'\omega\frac{\partial}{\partial \omega}
		-\bar{\xi}'\bar{\omega}\frac{\partial}{\partial \bar{\omega}}
		-\Big(
			\xi' + \bar{\xi}' + 2l
		\Big) \frac{\partial}{\partial \phi}\\
		+\big(
			l+ia
		\big)\Phi_1	\frac{\partial}{\partial \Phi_{1}}
		+\big(
			l-ia\big)\bar{\Phi}_{1} 
			\frac{\partial}{\partial \bar{\Phi}_{1}}\ ,
\end{aligned}
\end{equation}
where $a,l\in \mathbb{R}$, function
\begin{equation}
	\xi(z) = \frac{1}{\sqrt{2}}\left(
		\xi^{1}
		+i\xi^{2}
	\right)
\end{equation}
is holomorphic by (\ref{eq:xi_constraint}) and $l\neq0$ only if $\Lambda = 0$.


\begin{thebibliography}{1}
	\bibitem{H} 
		H\'ajicek P 1974 Three remarks on axisymmetric stationary horizons \emph{Commun. Math. Phys.} \textbf{36} 305--320
	\bibitem{MI} 
		Moncrief V and Isenberg J 1983 Symmetries of cosmological Cauchy horizons \emph{Commun. Math. Phys.} \textbf{89} 387--413
	\bibitem{ABL} 
		Ashtekar A, Beetle C and Lewandowski J 2002 
Geometry of generic isolated horizons \emph{Class. Quant. Grav.} \textbf{19} 1195--1225 (arXiv:gr-qc/0111067 [gr-qc])
	\bibitem{LP2003} 
		Lewandowski J and Paw\l{}owski T 2003 Extremal isolated horizons: a local uniqueness theorem \emph{Class. Quant. Grav.} \textbf{20} 587--606 (arXiv:gr-qc/0208032 [gr-qc])
	\bibitem{LP2005} 
		Lewandowski J and Paw\l{}owski T 2005 Quasi-local rotating black holes in higher dimension: geometry \emph{Class. Quant. Grav.} \textbf{22} 1573--1598 (arXiv:gr-qc/0410146 [gr-qc])
	\bibitem{PLJ} 
		Paw\l{}owski T, Lewandowski J and Jezierski J 2004 Spacetimes foliated by Killing horizons \emph{Class. Quant. Grav.} \textbf{21} 1237--1251 (arXiv:gr-qc/0306107 [gr-qc])
	\bibitem{LSW} 
		Lewandowski J, Szereszewski A and Waluk P 2016 Spacetimes foliated by nonexpanding and Killing horizons: Higher dimension \emph{Phys. Rev.} D \textbf{94} 064018 (arXiv:1605.07038 [gr-qc])
	\bibitem{LS} 
		Lewandowski J and Szereszewski A 2019 Spacetimes foliated by nonexpanding null surfaces in the presence of a cosmological constant \emph{Phys. Rev.} D \textbf{100} 024049
	\bibitem{K} 
		Kundt W 1961 The Plane-fronted Gravitational Waves \emph{Z. Physik} \textbf{163} 77--86
	\bibitem{PO} 
		Podolsky J and Ortaggio M 2003 Explicit Kundt type $II$ and $N$ solutions as gravitational waves in various type $D$ and $O$ universes \emph{Class. Quantum Grav.} \textbf{20} 1685--1701
	\bibitem{KL1} 
		Kunduri H K and Lucietti J 2009 A classification of near-horizon geometries of extremal vacuum black holes \emph{J. Math. Phys.} \textbf{50} 082502 (arXiv:0806.2051 [hep-th])
	\bibitem{KL2} 
		Kunduri H K and Lucietti J 2013 Classification of Near-Horizon Geometries of Extremal Black Holes \emph{Living Rev. Rel.} \textbf{16} 8 (arXiv:1306.2517 [hep-th])
	\bibitem{CST} 
		Chru\'sciel P T, Szybka S J and Tod P 2018 Towards a classification of vacuum near-horizons geometries \emph{Class. Quant. Grav.} \textbf{35} 015002 (arXiv:1707.01118 [gr-qc])
	\bibitem{LL} 
		Li C and Lucietti J 2016 Transverse deformations of extreme horizons \emph{Class. Quant. Grav.} \textbf{33} 075015 (arXiv:1509.03469 [gr-qc])
	\bibitem{KLS} 
		Kolanowski M, Lewandowski J and Szereszewski A 2019 Extremal horizons stationary to the second order: New constraints \emph{Phys. Rev.} D \textbf{100} 104057
	\bibitem{CCH} 
		Chru\'sciel P T, Lopes Costa J and Heusler M 2012 Stationary Black Holes: Uniqueness and Beyond \emph{Living Rev. Rel.} \textbf{15} 7 (arXiv:1205.6112 [gr-qc])
	\bibitem{CRT} 
		Chru\'sciel P T, Reall H S and Tod P 2006 On non-existence of static vacuum black holes with degenerate components of the event horizon \emph{Class. Quant. Grav.} \textbf{23}, 549--554
	\bibitem{DRKLS} 
		Dobkowski-Ry\l{}ko D, Kami\'nski W, Lewandowski J and Szereszewski A 2018 The Near Horizon Geometry equation on compact 2-manifolds including the general solution for $g>0$ \emph{Phys. Lett.} B \textbf{785} 381--385
	\bibitem{BH} 
		Bardeen J M and Horowitz G T 1999 Extreme Kerr throat geometry: A vacuum analog of AdS$_2\times \text{S}^2$ \emph{Phys. Rev.} D \textbf{60} 104030 (arXiv:hep-th/9905099 [hep-th])
	\bibitem{S} 
		Strominger A 1998 Black hole entropy from near-horizon microstates \emph{JHEP} \textbf{02} 009 (arXiv:hep-th/9712251 [hep-th])
	\bibitem{WT}
		 Wu X-N and Tian Y 2009 Extremal isolated horizon/CFT correspondence \emph{Phys. Rev.} D \textbf{80} 024014
	\bibitem{DGS} 
		Dunajski M, Gutowski J and Sabra W 2017 Einstein–Weyl spaces and near-horizon geometry \emph{Class. Quant. Grav.} \textbf{34} 045009
	\bibitem{stephani_symmetries} 
		Stephani H 1989 \emph{Differential Equations: Their Solution Using Symmetries} ed Maccallum M (Cambridge: Cambridge University Press)
	\bibitem{vinogradov} 
		Bocharov A V, Chetverikov V N, Duzhin S V, Khor'kova N G,  Samokhin A V, Torkhov Y N and Verbovetsky A M 1999 \emph{Symmetries and Conservation Laws for Differential Equations of Mathematical Physics} ed Krasil'shchik I S and Vinogradov A M (Providence, RI: American Mathematical Society)
	\bibitem{olver_lie} 
		Olver P J 1993 \emph{Applications of Lie Groups to Differential Equations} (New York City NY: Springer-Verlag) 2\textsuperscript{nd} edition
\end{thebibliography}
\end{document}